\documentclass[preprint,number, authoryear,12pt]{article}
\usepackage[margin=3cm]{geometry}

\title{ \bf Nonabelian Kaluza-Klein Dyon  \rm}

\author{\bf  Hsiang-Lin Lou\thanks{~Email: {sllou@thu.edu.tw}}
          and   Pei-Hsuan Li \rm \thanks{~Email: {ludmila.phys@gmail.com}}\\
       Deparment of Physics, Tunghai University\\
       Taichung, 407, Taiwan}
\date{\today}

\begin{document}
\maketitle
\vspace{3cm}
\begin{abstract}
 We propose a new nonabelian Kaluza-Klein dyon solution in
 the seven-dimen-sional
spacetimes. This is a Wu-Yang-like Kaluza-Klein dyon.
The dyonic metric is spherically symmetric, two spherical coordinate systems are used,
 and the metric
admits $SO(3)$ Killing vectors.  It can be verified
that the Einstein equation is satisfied in seven-dimensional spacetimes.
 The stress-energy tensor of the
Yang-Mills field in Einstein equation is derived from Ricci tensor automatically,
not put by hand from outside. The Ricci scalar curvature $\bar{R}$ vanishes.
The four-dimensional part of the dyonic metric is just the Schwarzschild black hole metric.

\end{abstract}
\vspace{1cm} \noindent PACS numbers: 04.50.Cd

\vspace{6cm}
\section{Introduction}
The well-known Gross-Perry-Sorkin solution
[1]
[2]
 in five-dimensional Kaluza-Klein (KK)  theory
is the simplest KK magnetic monopole
 solution associated with $U(1)$ abelian group.
 It has long been puzzled that whether the nonabelian   KK  Wu-Yang-like [3] monopole [4] or
 dyon solutions may exist? The Wu-Yang monopole means that it is a spherically symmetric
 point-like monopole without
 Dirac string.
 In this paper we will present a nonabelian   Wu-Yang-like  KK dyon solution
  in
 the seven-dimensional spacetimes.
  By calculating the Christoffel symbols and  the  Ricci tensor, it can be shown
that the Einstein equation is
 satisfied.
 In the next paper, we will show an alternative method by using the orthonormal frame
 and Cartan's structure equations. The Ricci tensor in that paper can be obtained via the affine spin
 connection one-form. The results from these two these different methods are coincident.

Suppose the  line element of the Kaluza-Klein theory [5]  in $(4+N)$ dimensions can be written as
  \begin{eqnarray}
  d \bar{s}^2 &=&\bar{g}_{AB}d\bar{x}^A d\bar{x}^B  \\
  &=&g_{\mu \nu} (x) dx^{\mu} dx^{\nu}    +
  \gamma_{mn}(y) (dy^m + B_{\mu}^m dx^{\mu} )(dy^n + B_{\nu}^n dx^{\nu} ),
  \end{eqnarray}
 where $x$ parametrizes   four-dimensional spacetimes, $y$ parametrizes extra dimensions.
 We use $A,B, C...$ indices to represent the total spacetimes; $\mu , \nu , \rho...$ to represent the four-dimensional
 spacetimes; $m, n, l...$ to represent the extra dimensions. $g_{\mu \nu}$ is only a function
 of $x$, and $\gamma_{mn}$ is only a function of $y$. The metric tensor is
 \begin{eqnarray}
 \bar{g}_{AB} &=& \left( \begin{array} {cc}
                       \bar{g}_{\mu \nu} & \bar{g}_{\mu n} \\
                        \bar{g}_{m \nu} & \bar{g}_{m n}
                        \end{array} \right)   \\
               &=& \left( \begin{array} {cc}
                         g_{\mu \nu}+ \gamma_{mn} B_{\mu}^m B_{\nu}^n
                          & B_{\mu}^m \gamma_{m n}\\
                         \gamma_{m n}B_{\nu}^n & \gamma_{m n}
                        \end{array} \right),
 \end{eqnarray}
while the inverse metric tensor is
\begin{eqnarray}
 \bar{g}^{AB} &=& \left( \begin{array} {cc}
                       \bar{g}^{\mu \nu} & \bar{g}^{\mu n} \\
                        \bar{g}^{m \nu} & \bar{g}^{m n}
                        \end{array} \right) \\
               &=& \left( \begin{array} {cc}
                         g^{\mu \nu}
                          & - g^{\mu \nu }B_{\nu}^n\\
                         -B_{\lambda }^m g^{\lambda \nu}
                         & \gamma^{m n}+ B_{\lambda}^m B_{\sigma}^n g^{\lambda \sigma}
                        \end{array} \right).
 \end{eqnarray}
$B_{\mu}^m $ cannot be identified as the Yang-Mills field. To extract the true Yang-Mills field,
one has to introduce the Killing vectors
\begin{equation}
L_a \equiv - i \zeta_{a}^{m} \partial_{m},
 \end{equation}
which generate a Lie algebra,
\begin{equation}
[L_{a} , L_{b} ]=i f_{ab}^{c} L_{c},
 \end{equation}
associated with some symmetry.  $a,b,c...$ are the gauge group indices.
$f_{ab}^{c}$ are the structure constants of a Lie algebra.
There is no need to differentiate between the lower and upper indices
for the gauge groups.
 Inserting $L_{a}$ of (7)
into equation (8), one gets the Killing's equation
\begin{equation}
\zeta_{a}^m \partial_{m} \zeta_{b}^n -
\zeta_{b}^m \partial_{m} \zeta_{a}^n
=- f_{ab}^{c} \zeta_{c}^n .
 \end{equation}
 With these Killing vectors, one can define
 \begin{equation}
B_{\mu}^m =\zeta_{a}^{m} A_{\mu}^{a},
 \end{equation}
where $A_{\mu}^{a}$ is the true Yang-Mills field and
$\zeta_{a}^{m}$ is only a function of $y$.

The Christoffel symbols defined by the form,
 \begin{equation}
\bar{\Gamma}_{AB}^{C}= {1\over 2} \;  \bar{g}^{CD}
( \partial_{A} \; \bar{g}_{BD} +
\; \partial_{B} \; \bar{g}_{AD} -
\; \partial_{D} \; \bar{g}_{AB} ),
 \end{equation}
can be given more explicitly by
\begin{eqnarray}
\bar{\Gamma}_{\mu \nu}^{\alpha}& =&  \Gamma_{\mu \nu}^{\alpha}
+ {1\over 2} g^{\alpha \lambda} \gamma_{mn} ( B_{\nu}^{m}
\widetilde{\mathcal{F}}_{\mu \lambda}^{n} +
B_{\mu}^{m} \widetilde{\mathcal{F}}_{\nu \lambda}^{n} )
+ {1\over 2} g^{\alpha \lambda} \gamma_{mn}
 ( B_{\mu}^{m} B_{\nu}^{l} +
B_{\nu}^{m} B_{\mu}^{l} ) \partial_{l} B_{\lambda}^{n}  \nonumber \\
& &\; \; \; \; \; \; + \; {1\over 2} g^{\alpha \lambda}
 B_{\mu}^{m} B_{\nu}^{n} B_{\lambda}^{l} \partial_{l} \gamma_{mn},
 \\
\bar{\Gamma}_{\mu \nu}^{m}& =&
-B_{\lambda}^{m}\bar{\Gamma}_{\mu \nu}^{\lambda} +
{1\over 2} ( \partial_{\mu} B_{\nu}^{m} +\partial_{\nu} B_{\mu}^{m} )
-{1\over 2}\gamma^{mn} \partial_{n} (B_{\mu}^{s} B_{\nu}^{t} \gamma_{st}),  \\
 \bar{\Gamma}_{m \nu}^{\alpha}&=& {1\over 2} g^{\alpha \lambda}
 \gamma_{ms} \widetilde{\mathcal{F}}_{\nu \lambda}^{s}
 + {1\over 2} g^{\alpha \lambda} B_{\lambda}^{l} B_{\nu}^{s}(\partial_{l} \gamma_{ms} )
    +{1\over 2} g^{\alpha \lambda}  B_{\nu}^{l}
 (\gamma_{ms} \partial_{l} B_{\lambda}^{s} + \gamma_{ls} \partial_{m} B_{\lambda}^{s}), \\
\bar{\Gamma}_{m n}^{\mu} &=& {1\over 2} g^{\mu \nu}\gamma_{nl}
\partial_{m} B_{\nu}^{l} + {1\over 2} g^{\mu \nu}\gamma_{ml}
\partial_{n} B_{\nu}^{l}
+ {1\over 2} g^{\mu \nu} B_{\nu}^{l}\partial_{l}\gamma_{mn}, \\
 \bar{\Gamma}_{n \nu}^{m} &=& -B_{\lambda}^{m}\bar{\Gamma}_{n \nu}^{\lambda}
 + {1\over 2} \gamma^{ml} \partial_{n}( B_{\nu}^{s} \gamma_{sl} )
 -{1\over 2} \gamma^{ml} \partial_{l}( B_{\nu}^{s} \gamma_{sn} ), \\
\bar{\Gamma}_{nl}^{m}& =& \widetilde{\Gamma}_{nl}^{m}
- {1\over 2}B_{\lambda}^{m} g^{\lambda \mu}(\gamma_{ls} \partial_{n}
B_{\mu}^{s} +\gamma_{ns} \partial_{l} B_{\mu}^{s})
- {1\over 2}B_{\lambda}^{m} g^{\lambda \mu}B_{\mu}^{s}
\partial_{s}\gamma_{nl},
\end{eqnarray}
where
\begin{eqnarray}
\widetilde{\mathcal{F}}_{\mu \nu}^{m} &\equiv&
\partial_{\mu} B_{\nu}^{m} + B_{\nu}^{l} \partial_{l} B_{\mu}^{m}
-(\mu \leftrightarrow \nu ) \\
&=& \zeta_{a}^{m} F_{\mu \nu}^{a}.
\end{eqnarray}
 $F_{\mu \nu}^{a}$  is the true field strength tensor of the
Yang-Mills field,
\begin{equation}
F_{\mu \nu}^{a} = \partial_{\mu} A_{\nu}^{a}- \partial_{\nu} A_{\mu}^{a}
+f_{bc}^{a} A_{\mu}^{b} A_{\nu}^{c}.
\end{equation}
$\Gamma_{\mu \nu}^{\alpha}$
in (12) are the  Christoffel symbols of the four-dimensional
spacetimes, $\Gamma_{\mu \nu}^{\alpha}
={1\over2} g^{\alpha \lambda}(\partial_{\mu} g_{\nu \lambda}
+\partial_{\nu} g_{\mu \lambda}
-\partial_{\lambda} g_{\mu \nu}) $ , while
$\widetilde{\Gamma}_{nl}^{m}$ in (17) are the  Christoffel symbols of the extra dimensions,
$\widetilde{\Gamma}_{nl}^{m}
={1\over2} \gamma^{ms }(\partial_{n} \gamma_{ls}
+\partial_{l} \gamma_{ns}
-\partial_{s} \gamma_{nl}) $.

\section{SO(3) Kaluza-Klein Dyon}
Now let us consider an ans$\ddot{a}$tz  of the  Kaluza-Klein dyonic metric admitting $SO(3)$
Killing vectors. The metric
is spherically symmetric, not only for the four-dimensional spacetimes but also for the extra dimensions.
 The line element can be written as the form
\begin{eqnarray}
d\bar{s}^2 =& -& e^{2\Psi} dt^{2} \nonumber \\
 &+& e^{2\Lambda} dr^{2} +r^2 d\theta^2 +r^2 \sin^{2}{\theta} d\phi^2
\nonumber \\
&+& e^{2\chi}(dR + B_{\mu}^{5} dx^{\mu} )^2
+ R^2 (d\Theta  + B_{\mu}^{6} dx^{\mu} )^2
+ R^2 \sin^{2}\Theta (d\Phi  + B_{\mu}^{7} dx^{\mu} )^2 .
\end{eqnarray}
$r,\theta , \phi $ are  three coordinates of the  ordinary   three-dimensional spherical coordinate system,
$(r,\theta ,\phi )=(\bar{x}^1 ,\bar{x}^2 , \bar{x}^3 )=(x^1 ,x^2 , x^3 )$.
 $R, \Theta , \Phi$
are another three coordinates of  the spherical coordinate system in the extra dimensions,
 $(R, \Theta ,\Phi ) =(\bar{x}^5 ,\bar{x}^6 , \bar{x}^7 )=(y^{5} ,y^6 ,y^7 )$.
$\Psi$ and $\Lambda$ are two functions of $r$, while $\chi $ is only a function of $R$.
 We have $g_{00}=-e^{2\Psi}$, $g_{11}=e^{2\Lambda}$, $g_{22} =r^2$, $g_{33}=r^2 \sin^2 \theta$, $\gamma_{55}=e^{2\chi}$,
 $\gamma_{66}=R^2$, $\gamma_{77}=R^2 \sin^2 \Theta$.

The gauge-field components of  the Wu-Yang-like KK dyon   are
\begin{eqnarray}
A_{0}^{a} &=& {1\over r} \hat{r}^{a}, \;\; \; \;\;\; \; \;
A_{1}^{a}  = 0, \\
A_{2}^{a} &=& - \hat{\phi}^{a}, \; \;\; \; \; \;\;
A_{3}^{a}  =  \sin\theta  \; \hat{\theta}^{a} .
\end{eqnarray}
$A_{1}^{a}$,$A_{2}^{a}$,$A_{3}^{a}$ are just the spherical coordinate representation of the Wu-Yang
monopole field in the ordinary gauge theory of four-dimensional spacetimes. The electric field of the
KK  dyon is
\begin{equation}
F_{01}^{a} = {1\over r^2} \;  \hat{r}^{a} ,
\end{equation}
while the magnetic field  is
\begin{equation}
F_{23}^{a} = - \sin \theta  \; \hat{r}^{a}.
\end{equation}

Since the relevant  Killing vectors are
\begin{eqnarray}
L_1 &=& -i ( -\sin\Phi {\partial \over {\partial \Theta }} - \cot \Theta \cos \Phi
{\partial \over {\partial \Phi}} ), \\
L_{2} &=& -i  ( \cos\Phi {\partial \over {\partial \Theta }} - \cot \Theta \sin \Phi
{\partial \over {\partial \Phi}} ), \\
L_{3} &=& -i ({\partial \over {\partial \Phi}} ),
\end{eqnarray}
which are just the generators of the $SO(3)$  group, then one has
 \begin{eqnarray}
\zeta_{a}^{5} &=& 0, \\
\zeta_{a}^{6} &=& \hat{\Phi}_{a}, \\
\zeta_{a}^{7} &=& -{1\over \sin{\Theta} } \; \hat{\Theta}_{a}.
\end{eqnarray}
$\hat{R}_{a}$,
$\hat{\Theta}_{a} $,
$ \hat{\Phi}_{a} $ are three unit vectors of the spherical coordinate system
of the extra dimensions.
It  can be checked that above three equations of $\zeta_{a}^{m}$ also
satisfy the Killing equation (9). The fields $B_{\mu}^{m}$ in (10) can be rewritten as
\begin{eqnarray}
B_{\mu}^{5} &=& \; 0,    \; \; \; \; \; \; \; \; \; \; \; \; \; \;  \; \; \; \; \; \;
B_{1}^{m} =\; \; 0,\; \; \; \; \; \; \; \; \; \; \; \;  \; \; \; \; \;  \\
B_{0}^{6}&=& {1\over r} \; \hat{r}  \cdot  \hat{\Phi}, \; \; \; \; \; \; \; \; \; \; \; \; \; \;
B_{0}^{7} = - { 1\over  {r\sin\Theta}} \; \hat{r} \cdot  \hat{\Theta}, \\
B_{2}^{6}&=&   - \hat{\phi}  \cdot  \hat{\Phi}, \; \; \; \; \; \; \; \; \; \; \; \;\; \;
B_{2}^{7} = \; \;\;  { 1\over  {\sin\Theta}} \; \hat{\phi} \cdot  \hat{\Theta}, \\
B_{3}^{6}&=&  \; \sin\theta \;  \hat{\theta}  \cdot  \hat{\Phi}, \; \; \; \; \; \; \; \; \;
B_{3}^{7} = - {\sin\theta \over  {\sin\Theta}} \; \hat{\theta} \cdot  \hat{\Theta}.
\end{eqnarray}
The nonzero components of $\widetilde{\mathcal{F}}_{\mu \nu}^{m}$ are
 \begin{eqnarray}
 \widetilde{\mathcal{F}}_{01}^{6} &=& \; \; {1\over r^2}\;  \hat{r} \cdot  \hat{\Phi},
\;\; \; \; \; \; \; \; \; \; \; \; \; \; \;
\widetilde{\mathcal{F}}_{01}^{7} = {-1 \over r^2\sin\Theta}\;  \hat{r} \cdot  \hat{\Theta},
\; \; \; \; \; \; \; \\
\widetilde{\mathcal{F}}_{23}^{6} &=&  {- \sin\theta}\;  \hat{r} \cdot  \hat{\Phi},
\;\; \; \; \; \; \; \; \; \; \;
\widetilde{\mathcal{F}}_{23}^{7} = {\sin\theta \over \sin\Theta}\;  \hat{r} \cdot  \hat{\Theta}.
\; \; \; \; \; \; \;
\end{eqnarray}

Because of the chosen dyonic metric, there exist the following identities:
 \begin{eqnarray}
& &\gamma_{77} \; \partial_{6} B_{\mu}^{7} + \gamma_{66} \; \partial_{7} B_{\mu}^{6} =0,
\; \; \; \;\; \; \; \; \; \; \; \; \; \; \; \; \;\\
& &\gamma_{77} \; \partial_{7} B_{\mu}^{7} + {1\over 2}
 B_{\mu}^{6} \; \partial_{6} \gamma_{77}   =0,
\end{eqnarray}
which simplify our calculations drastically.
 For example, from (12) and (14), we get the simplified
Christoffel symbols,
\begin{eqnarray}
\bar{\Gamma}_{\mu \nu}^{\alpha} &=&  \Gamma_{\mu \nu}^{\alpha}
+ {1\over 2} g^{\alpha \lambda} \gamma_{mn} ( B_{\nu}^{m}
\widetilde{\mathcal{F}}_{\mu \lambda}^{n} +
B_{\mu}^{m} \widetilde{\mathcal{F}}_{\nu \lambda}^{n} ), \\
 \bar{\Gamma}_{m \nu}^{\alpha} &=& {1\over 2} g^{\alpha \lambda}
 \gamma_{ms} \widetilde{\mathcal{F}}_{\nu \lambda}^{s}.
\end{eqnarray}

Among
$196 \; (=7\times 28 )$ independent components of the Christoffel symbols,
 there are 82 independent
components
are  nonzero.
Substituting these into  the
 Ricci tensor,
\begin{equation}
\bar{R}_{BD} =\partial_{A} \bar\Gamma_{BD}^{A} - \partial_{D} \bar\Gamma_{BA}^{A}
+\bar{\Gamma}_{AE}^{A}\bar{\Gamma}_{BD}^{E}
 -\bar{\Gamma}_{DE}^{A}\bar{\Gamma}_{AB}^{E} .
\end{equation}

During the  lengthy calculations,
one will use frequently the following identities:
\begin{eqnarray}
\partial_{\theta} \hat{r} &=& \hat{\theta}, \; \; \; \; \; \;  \; \; \; \; \; \;
 \; \; \; \;   \; \; \; \;  \; \; \; \; \; \;
\partial_{\Theta} \hat{R} = \hat{\Theta}, \; \; \; \;  \; \; \; \;  \; \; \; \;\; \;
 \; \; \; \;  \; \;     \\
\partial_{\theta} \hat{\theta} &=& -\hat{r}\; \; \; \; \; \;  \; \; \; \; \; \;
 \; \; \; \;   \; \; \; \;  \; \; \;\; \;
 \partial_{\Theta} \hat{\Theta} = -\hat{R},\; \; \; \;  \; \; \; \;  \; \; \; \;
 \; \; \; \;  \; \;    \\
\partial_{\phi} \hat{r} &=& \sin\theta \; \hat{\phi}\; \; \; \; \; \;  \; \; \; \; \; \;
 \; \; \; \;   \; \; \;
 \partial_{\Phi} \hat{R} = \sin\Theta \; \hat{\Phi},\; \; \; \;  \; \; \; \;  \; \; \; \;
 \; \; \; \;  \; \;    \\
 \partial_{\phi} \hat{\theta} &=& \cos\theta \; \hat{\phi}\; \; \; \; \; \;  \; \; \; \; \; \;
 \; \; \; \;   \; \; \;
 \partial_{\Phi} \hat{\Theta} = \cos\Theta \; \hat{\Phi},\; \; \; \;  \; \; \; \;  \; \; \; \;
 \; \; \; \;  \; \; \; \;    \\
 \partial_{\phi} \hat{\phi} &=& - \sin\theta \; \hat{r} -\cos\theta \; \hat{\theta} ,\\
 \partial_{\Phi} \hat{\Phi} &=& - \sin\Theta \; \hat{R} -\cos\Theta \; \hat{\Theta} ,
\end{eqnarray}
\begin{eqnarray}
\hat{r}_{a} \hat{r}_{b} +\hat{\theta}_{a} \hat{\theta}_{b} +
\hat{\phi}_{a} \hat{\phi}_{b} &=& \delta_{ab}, \\
\hat{R}_{a} \hat{R}_{b} +\hat{\Theta}_{a} \hat{\Theta}_{b} +
\hat{\Phi}_{a} \hat{\Phi}_{b} &=& \delta_{ab}.
\end{eqnarray}
\begin{eqnarray}
(\hat{\theta} \cdot \hat{\Theta}) \; (\hat{\phi} \cdot \hat{\Phi}) -
(\hat{\theta} \cdot \hat{\Phi}) \; (\hat{\phi} \cdot \hat{\Theta})
&=&  \hat{r} \cdot \hat{R} \\
(\hat{r} \cdot \hat{\Theta}) \; (\hat{\theta} \cdot \hat{\Phi}) -
(\hat{r} \cdot \hat{\Phi}) \; (\hat{\theta} \cdot \hat{\Theta})
&=&  \hat{\phi} \cdot \hat{R}
\end{eqnarray}

 After many miracle cancellations, it can be found  that the components of the dyonic metric
 are
 \begin{eqnarray}
g_{00} &=& - (1-{r_{s} \over r}) , \\
g_{11} &=&  \; \; \;  (1-{r_{s} \over r })^{-1},\\
\gamma_{55} &=& \; \; \; 1,
\end{eqnarray}
where $r_s$ is the Schwarzschild radius.
Then $\bar{R}_{mn} $, $\bar{R}_{\mu m} $  and almost $\bar{R}_{\mu \nu} $  are
equal to zero except
 \begin{eqnarray}
\bar{R}_{00}& =& -{1\over 2} g^{11} {R^2 \over r^4} \{ ( \hat{r} \cdot \hat{\Theta} )^2
+( \hat{r} \cdot \hat{\Phi} )^2 \}, \\
\bar{R}_{11}& =& -{1\over 2} g^{00} {R^2 \over r^4} \{ ( \hat{r} \cdot \hat{\Theta} )^2
+( \hat{r} \cdot \hat{\Phi} )^2 \} ,\\
\bar{R}_{22}& =& - {R^2 \over {2 r^2}} \{ ( \hat{r} \cdot \hat{\Theta} )^2
+( \hat{r} \cdot \hat{\Phi} )^2 \}, \\
\bar{R}_{33}& =& - {{R^2 \sin^2 \theta} \over {2 r^2}} \{ ( \hat{r} \cdot \hat{\Theta} )^2
+( \hat{r} \cdot \hat{\Phi} )^2 \}.
\end{eqnarray}
 So one has the Ricci scalar curvature,
\begin{eqnarray}
\bar{R} &=& \bar{g}^{AB} \bar{R}_{AB}  \\
&=& g^{00} \bar{R}_{00} +g^{11} \bar{R}_{11}
+ g^{22} \bar{R}_{22} +g^{33} \bar{R}_{33} \\
&=& 0.
 \end{eqnarray}
From the fields $\widetilde{\mathcal{F}}_{\mu \nu}^{m}$ in
(36) and (37), the components of the  Ricci tensor, $(56) \sim (59)$, can be recast into the form,
\begin{equation}
\bar{R}_{\mu \nu} = -{1\over 2}\bar{g}^{\alpha \beta}  \gamma_{mn}
\widetilde{\mathcal{F}}_{\mu \alpha}^{m}
\widetilde{\mathcal{F}}_{\nu \beta}^{n}
\end{equation}
Since the identity,
$  \gamma_{mn}
\widetilde{\mathcal{F}}_{\mu \nu}^{m}
\widetilde{\mathcal{F}}^{\mu \nu n} =0$, holds, the right-hand side of the equation (63) can be
identified as $8\pi$ times
the stress-energy tensor of the Yang-Mills field,
\begin{equation}
\bar{R}_{\mu \nu} = 8\pi \bar{T}_{\mu \nu}, \; \; \; \;
\bar{T}_{\mu \nu} ={-1\over 16 \pi}\bar{g}^{\alpha \beta}  \gamma_{mn}
\widetilde{\mathcal{F}}_{\mu \alpha}^{m}
\widetilde{\mathcal{F}}_{\nu \beta}^{n}.
\end{equation}
 Then the Einstein equation, $\bar{R}_{AB}-{1\over 2}
 \bar{g}_{AB} \bar{R} = 8\pi \bar{T}_{AB} $, is satisfied, where some components of $
 \bar{T}_{AB}$ are zero,
 $\bar{T}_{\mu m}=0$
 and $\bar{T}_{ m n}=0.$

\section{Discussions}
We have shown that the  $ SO(3)$  dyon solution satisfies the  Einstein equation in
 the seven-dimensional
spacetimes.  The stress-energy tensor of the
Yang-Mills field in Einstein equation is derived from Ricci tensor automatically,
not put by hand from outside.
That the four-dimensional part of the KK dyonic metric is just  the Schwarzschild black hole metric
 means  the black hole in four-dimensional spacetimes
and the dyonic gauge field are intimately
related from the higher-dimensional spacetimes point of view.
Since the dyon solution is spherically symmetric and the associated nonabelian symmetry group is
$SO(3)$, many generalizations are possible. Applied  to string theories maybe the most important
direction for further investigations.

\end{document}